# Efficient Electric Field Control of Magnetic Phase in Bilayer Magnets via interlayer hopping modulation

B. Liu[1], J. S. Feng[2], H. J. Xiang[3,4,5]*, Z. Dai[1]*, and Zhi-Xin Guo[1,†]

[1]State Key Laboratory for Mechanical Behavior of Materials, Xi'an Jiaotong University, Xi'an, Shaanxi, 710049, China.

[2]School of Physics and Materials Engineering, Hefei Normal University, Hefei 230601, China.

[3]Key Laboratory of Computational Physical Sciences (Ministry of Education), Institute of Computational Physical Sciences, and Department of Physics, Fudan University, Shanghai 200433, China.

[4]Shanghai Qi Zhi Institution, Shanghai 200030, China.

[5]Collaborative Innovation Center of Advanced Microstructures, Nanjing 210093, China.

Authors to whom correspondence should be addressed: *hxiang@fudan.edu.cn; *sensdai@mail.xjtu.edu.cn; †zxguo08@xjtu.edu.cn

**Abstract:**

Two-dimensional (2D) van der Waals (vdW) magnets present a promising platform for spintronic applications due to their unique structural and electronic properties. The ability to electrostatically control their interlayer magnetic coupling between ferromagnetic and antiferromagnetic phases is particularly advantageous for the development of energy-efficient spintronic components. While effective in bilayer $CrI_3$, achieving this control in other 2D magnets remains a challenge. In this work, we demonstrate that bilayer $Cr_2Ge_2Te_6$ can achieve efficient electrostatic control through interlayer hopping modulation. We show that an external electric field can effectively manipulate the FM↔AFM phase transition when interlayer hopping is enhanced by pressure or sliding. We further develop a four-site interlayer hopping model, revealing that the phase transition is driven by a combined effect of on-site energy splitting and interlayer electronic hopping. These findings pave the way for designing novel, electrically tunable spintronic devices, offering substantial potential for energy-efficient information processing and storage.

The discovery of spintronic phenomena, such as tunneling magnetoresistance (TMR) [1], has opened new avenues for controlling the spin and charge properties of materials, thereby enabling the development of innovative functional spintronic devices [2-7]. Two-dimensional (2D) van der Waals (vdW) magnetic material, such as $CrX_3$(X=Cl, Br, I) [8-10], $CrYTe_3$ (Y=Ge, Si) [11,12] and $Fe_mGeTe_2$ (m=3, 5) [13,14] etc, are among the most promising candidates for these devices. In contrast to the traditional bulk materials, 2D materials can achieve atomic thickness and their interfaces exhibit atomic-level smoothness. Although these materials interact weakly through interlayer vdW forces, the magnetism is sensitive to interlayer coupling. It had been found that the interlayer magnetic coupling can be efficiently modulated by sliding, rotation, and applying pressure [15-18]. However, the manipulation of the interlayer phase transition via an external electronic field has been rarely reported [19].

The application of an external electric field for magnetism control shows significant promise in advancing high-speed, low-power spintronic devices, as it does not require the involvement of spin or charge currents, resulting in nearly zero energy dissipation [20,21]. This has garnered significant attention in the field of spintronics [22,23]. However, as far as we know, except for bilayer $CrI_3$ [24-27], achieving successful control of ferromagnetic (FM) ↔ antiferromagnetic (AFM) phase transition by an external electric field remains a challenge in other 2D magnetic materials.

In this work, by using the first-principles calculations, we discover that an efficient electrostatic control of bilayer $Cr_2Ge_2Te_6$ (CGT) can be generally achieved in with modulation of interlayer hopping. In bilayer CGT, we demonstrate that the FM↔AFM phase transition of a 2D material can be effectively manipulated by an external electric field, when the interlayer hopping is enhanced via pressure or sliding. We also construct a four-site interlayer hopping model to describe the FM↔AFM phase transition, revealing the role of interlayer electronic hopping induced by pressure/sliding and energy splitting induced by the external electric field on the phase transition.

We first show that the magnetic ground state of 2D vdW material strongly correlates to the interlayer coupling strength, characterized by the interlayer electronic hopping. An efficient approach of modulating the interlayer hopping is changing the interlayer distance (d), achievable via applying external pressure. The calculation method is shown in Supplemental Material. Our computational findings reveal a distinct correlation between the interlayer distance and the magnetic ground state in the original bilayer CGT with AB stacking order (denoted as AB CGT). The structure of AB CGT is depicted in Figs. 1(a) and 1(b), where Cr atoms are positioned within the octahedral crystal field created by Te atoms, consistent with the structure obtained experimentally [11]. The ground state of AB CGT with the original layer spacing (3.26 Å) exhibits FM exchange coupling of both intralayer and interlayer, consistent with previous studies [11]. As shown in Fig. 1(e), as the interlayer distance decreases, the CGT bilayer undergoes an interlayer magnetic phase transition from a FM ground state to an AFM state, notably occurring when the interlayer spacing decreases below 2.90 Å. Conversely, as the layer spacing increases, with the decrease in ΔE evident ($\Delta E = E_{AFM} - E_{FM}$, where $E_{AFM}$ and $E_{FM}$ are the total energy of CGT in interlayer AFM and FM state, respectively), the ground state remains unchanged.

Another effective way to modulate the interlayer coupling strength is to change the stacking order between the two layers, which can be achieved by interlayer sliding. Hence, we further calculate the variation of ΔE with interlayer sliding of bilayer CGT, from an AB stacking (the most stable structure, Fig. 1(a)) to an AA stacking structure (Fig. 1(c), its structural stability is confirmed by the absence of imaginary phonon frequencies shown in Fig. S1(b)). The full sliding path is additionally shown in Fig. S1(c). As depicted in Fig. 1(f), FM ground state is exclusively observed for the AB structure, whereas AFM state emerges under all other stacking sequences. Note that the energy barrier of bilayer CGT is 74 meV per Cr (Fig. S1(a)), aligning with the barrier observed for the triple-layer 3R-$MoS_2$ structure [28], implying the feasibility of realizing interlayer sliding in bilayer CGT. Subsequent discussion reveals that either the decrease of interlayer distance or the sliding from AB structure to AA structure has an effect of enhancing the electron hopping, which makes the ground state favor the interlayer AFM state.

Next, we discuss the effect of external electric field on the magnetic phase transition of bilayer CGT, which can be characterized by the on-site energy splitting. As shown in Fig. 1(f), the ground state of AB CGT without external pressure (d = 3.26 Å) and electric field is the FM state. As the electric field (E) increases, ΔE monotonically increases, thereby stabilizing the FM state. This outcome highlights the advantageous role of energy splitting in promoting FM interlayer coupling. Given the enhancement of interlayer coupling, which forms the AFM interlayer state, efficient manipulation on the FM↔AFM phase transition is expected under the effect of interlayer electron hopping and energy splitting.

To confirm above argument, we conduct additional calculations on the variation in energy difference between AFM and FM states of bilayer CGT with an external electric field, under smaller interlayer distance (2.90 Å) and a different stacking structure (AA). As shown in Fig. 1(f), in both cases, although the ground state is AFM (ΔE<0) without electric field, ΔE remarkably increases with the electric field, leading to an AFM to FM phase transition. The difference is that when AA stacking structure, the critical electric field is greater than 0.9 V/Å, while in AB-2.90 Å the critical electric field is greater than 0.7 V/Å. When considering spin-orbit coupling (SOC), the critical electric field becomes about 0.7 V/Å as shown in Fig. S2. The results clearly show that a precise control of interlayer FM↔AFM phase transition can be achieved by enhancing the interlayer electron hopping. From the viewpoint of practical applications, one only needs to fix a bilayer CGT to a certain interlayer distance via external pressure or a certain stacking structure via sliding by control-sliding-orders-engineering-approaches [29], then the modulation of FM↔AFM phase transition can be effectively realized by applying an external electric filed.

Then we come to discuss how the interlayer electron hopping and energy splitting are affected by the interlayer distance/stacking structure and the external electric field. Based on structural analysis, as shown in Fig. S3, the nearest neighbor (NN) spin exchange interaction between layers is primarily facilitated by super-superexchange pathways involving Cr-Te-Te-Cr linkages. This means that the transition of NN spin exchange, produced by decreasing interlayer and sliding, is mainly caused by the change of super-superexchange. To reveal the underlying mechanism, we further explore the two effects on the hoping of Cr-Te-Te-Cr linkages. Table S1 shows the calculated bond lengths of intralayer Cr-Te and interlayer Te-Te for the AB structure with d=2.90 Å and AA structure obtained by sliding (d=3.33 Å), in comparison with that of AB structure with d=3.26 Å. It is found that both decreasing interlayer distance and sliding hardly change the intralayer Cr-Te bond length (Δd≤0.01 Å). Nonetheless, they have prominent influence on the interlayer Te-Te bond length, which is reduced by 0.28 Å and 0.12 Å for the AB structure with d=2.90 Å and AA structure, respectively. The smaller interlayer distance corresponds to stronger interlayer hopping, which can be verified by the interlayer orbital overlaps, dominated by the p-orbital of Te atoms between layers (see Fig. S4 for details). As depicted in Figs. 2(a)-2(c), compared to AB CGT with d = 3.26 Å, the overlap of interlayer p-orbitals significantly increases in case of either decreasing interlayer distance or sliding to AA structure. Since $E_{AFM}-E_{FM}$ is very sensitive to the interlayer distance and stacking structure (Fig. 1), this result shows that the interlayer electron hopping plays a pivotal role in determining the interlayer spin exchange of bilayer CGT. Note that, as the interlayer hopping increases, the interlayer spin exchange transforms from a FM coupling to an AFM coupling.

On the other hand, the external electric field generally has a significant effect on the on-site energy. Hence, a vertical electric field can induce energy band splitting in a bilayer material [26,27,30,31]. Figs. 2 (d) and 2(e) show the calculated partial density of states (PDOS) of CGT (AB stacking with d=3.26 Å, in AFM state) for the Cr atoms and Te atoms at the interface of the bilayer structure. Notably, the vertical electric field leads to a significant split in the energy levels of both the d orbital of Cr and p orbital of Te between the two CGT layers. This result is in line with prior researches, where the energy bands exhibit a systematic departure from degeneracy as the applied electric field strength increases [26,27,30,31]. We have also explored the effect of external electric field on the CGT with either AB (d=2.9 Å) or AA structure, and observed the sizable orbital-energy splitting in both cases. Fig. S5 additionally shows the electric-field-strength dependent splitting energy between $t_{2g}$ orbitals of the top and bottom layers with SOC (denoted as $\Delta_0$), for CGTs both with AB (d=2.9 Å) and AA structures. Notably, $\Delta_0$ increases linearly with E, regardless of distinct interlayer distance and stacking structure. This result indicates that the external electric field induced orbital-energy splitting has a profound influence on the interlayer spin exchange interaction, which can lead to an AFM to FM phase transition.

Based on the above analysis, we construct a tight binding Hamiltonian to reveal the underlying mechanism for FM↔AFM phase transition. As depicted in Fig. S6, the PDOS near the Fermi level is predominantly composed of Cr and Te elements, with a negligible contribution from Ge. This indicates that the hybridization primarily occurs between Cr-d and Te-p orbitals, which aligns with our previous analysis of super-superexchange. To effectively capture the essential physics, we implement appropriate simplifications. Essentially, considering their close energy levels, it is reasonable to use double-orbital representing two different manifolds ($t_{2g}$ and $e_g$ manifolds in Cr of CGT) and single orbital representing one manifold (p orbital manifold in Te of CGT). Meanwhile, the weak Cr-Ge interaction can be disregarded without compromising the accuracy of this study's findings. The interlayer hopping model illustration, as depicted in Fig. 3(a), considers both intralayer superexchange and interlayer super-superexchange, which are crucial for understanding the system's behavior. Consequently, the tight binding Hamiltonian for the bilayer CGT can be expressed as [32,33]:

$$H = \sum_{im\sigma} \epsilon_{im} \hat{a}^\dagger_{im\sigma} \hat{a}_{im\sigma} + \sum_{imjn\sigma} T_{im,jn} \hat{a}^\dagger_{im\sigma} \hat{a}_{jn\sigma} - U \sum_i \boldsymbol{m}_i \cdot \boldsymbol{S}_i. \qquad (1)$$

Here i, j mean the site indices, and m, n represent the orbital indices. $\epsilon$ and $T$ are the onsite energy and the hopping parameter between different orbitals, respectively. σ represents the spin (σ = ↑ and ↓ for spin up and down, respectively). $\boldsymbol{m}_i$ is the unit vector of the magnetic moment at the i-th site, and U is the onsite repulsion. $\boldsymbol{S}_i$ is the electronic spin operator at site i.

Figs. 3(b) and 3(c) further present schematic diagrams for the onsite energy and hopping parameters of the Hamiltonian for both FM and AFM states, under conditions with and without an external electric field. Note that, here we develop a four-site interlayer hopping model where the hopping interaction ($t_2$) of interlayer Te-Te atoms are taken. Distinct from the previous studies where a phenomenological description for the bilayer magnets has been adopted [27], this model can accurately describe superexchange interaction and give rise to a more comprehensive phase diagram for the bilayer magnets. The detailed Hamiltonians for FM and AFM states, with and without an external electric field, are shown in the Supplementary Material (Eqs. S1 and S2). Based on these

Hamiltonians, one can easily obtain the exchange energy $\Delta E = E_{FM} - E_{AFM}$ by employing the numerical approach (Supplementary Material S-IX).

Figure 4(a) shows the exchange energy as a function of interlayer hopping, calculated based on the four-site interlayer hopping model. It is seen that with increasing interlayer hoping $t_2$, $\Delta E$ initially increases slightly and then significantly decreases, becoming negative when $t_2 > 0.5$ eV. This indicates that an increase in $t_2$ can induce an FM to AFM phase transition in bilayer CGT. This result agrees well with our DFT calculations, demonstrating that the model effectively capture the correlation between interlayer magnetic coupling and interlayer electronic hopping. It is noteworthy that our four-site interlayer hopping model can be extended to other bilayer magnetic materials. For example, in transition metal dichalcogenide bilayers such as $CrX_2$ (X=S, Se, Te) [34] and $1T-VX_2$ (X=S, Se) [35], the interlayer-coupling-strength dependent ground magnetic phase can be naturally understood from the model.

Similarly, numerical results for $\Delta E$ can be also obtained when an external electric field is applied, resulting in a more complex FM↔AFM phase diagram (Fig. 4(b)) compared to that without an external electric field (Fig. 4(a)). As shown in Fig. 4(b), the FM and AFM phases coexist under varying $\Delta_0$ when $t_2$ is in range of [0.5, 0.7] eV, showing that the electric field can efficiently manipulate the ground state of CGT when the interlayer hoping is strong enough. This feature shows that the magnetic phase transition can be effectively modulated by an external electric field when the interlayer hopping is in certain strength, aligning with the above DFT calculations. Conversely, when $t_2$ is below 0.5 eV, with the external electric field increases, the magnetic ground state favors a FM state, aligning with the DFT calculations. Hence, the interlayer hopping model essentially captures relationship among the magnetic ground phase, interlayer hopping, and external electric filed for the bilayer magnets.

In summary, based on the DFT calculations, we have discovered an efficient way to achieve precise control of FM↔AFM phase transition in bilayer CGT by an external electric field, i.e., modulation of interlayer hopping via pressure, sliding, etc. This method capitalizes on the synergistic effects of interlayer electronic hopping and on-site energy splitting induced by electric field, thus enabling the desired phase transitions under an external electric field. We have also developed a four-site interlayer hopping model that integrates the effects of both the interlayer hopping and an external electric field. This model elucidates the underlying mechanisms governing the FM↔AFM phase transition control, offering insights into the nuanced interplay of magnetic states. Our study paves a way for the design of new electrically controllable spintronic devices with enhanced performance and potential applications in information processing and storage technologies.

## Acknowledgements:

This work was supported by the Ministry of Science and Technology of the People's Republic of China (Grant No. 2022YFA1402901), Natural Science Foundation of China (No. 12474237, 52371236), and Science Fund for Distinguished Young Scholars of Shaanxi Province (No. 2024JC-JCQN-09).

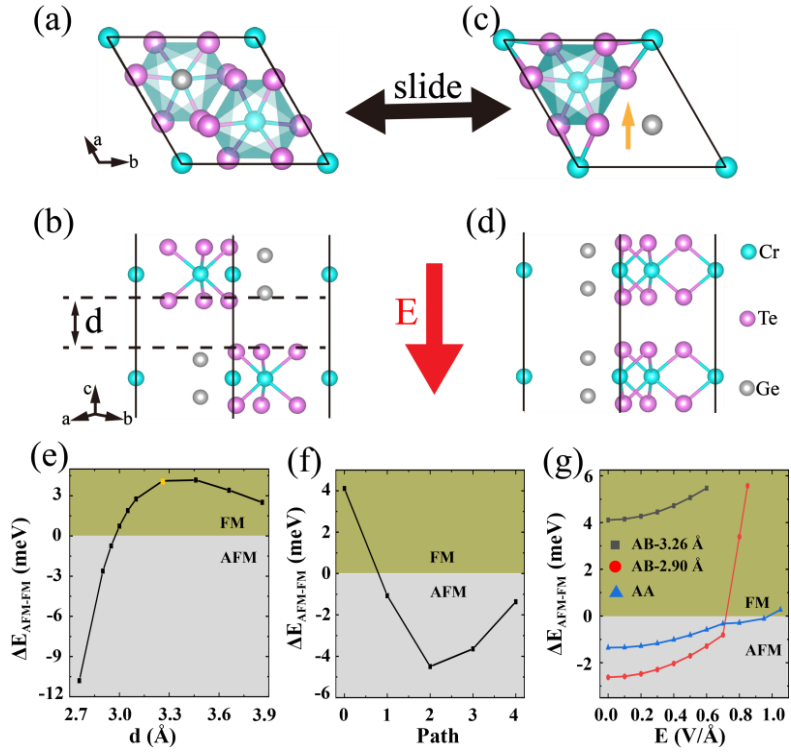

**Figure 1** (a) Top view and (b) side view of AB stacking bilayer CGT, with the definition of interlayer spacing d. (c) Top view and (d) side view of AA stacking bilayer CGT. Orange arrow and red arrow represents the direction of sliding and the external vertical electric field, respectively. (e) The energy difference between AFM and FM phase ($\Delta E = E_{AFM} - E_{FM}$) as a function of the interlayer distance d for bilayer AB stacking CGT. The orange square in (e) corresponds to case of the original layer spacing (d=3.26 Å). (f) The energy difference between AFM and FM phase as a function of sliding path from AB stacking structure (path-0) to AA stacking structure (path-4). (g) The energy difference between AFM and FM phase in the AB CGT with d=3.26 Å, AB CGT with d=2.90 Å and AA CGT as a function of the external electric field.

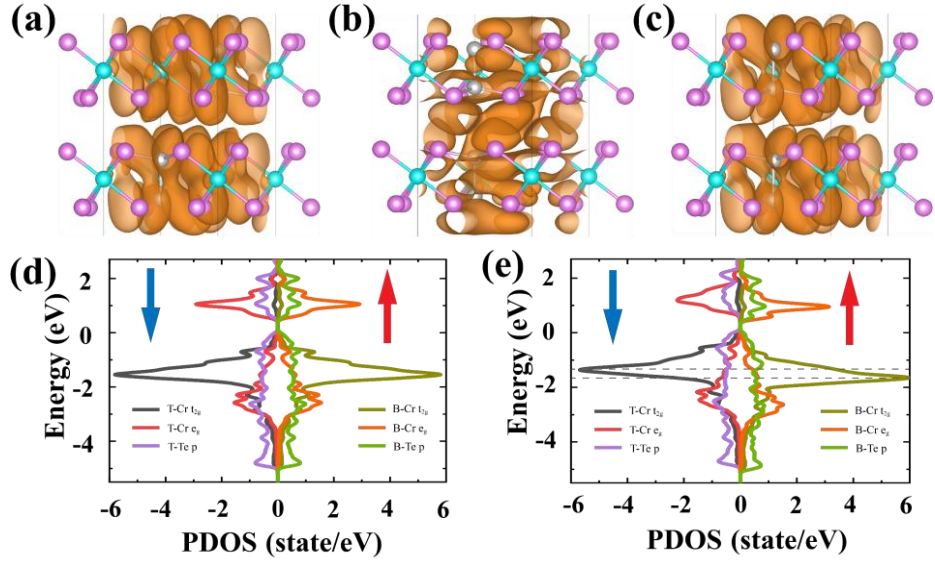

**Figure 2** (a)-(c) The squared wave-function of p-orbitals of across-layer Te atoms in FM state: (a) AB CGT with d=3.26 Å, (b) AB CGT with d=2.90 Å and (c) AA CGT, respectively. (d)-(e) The xxx (PDOS) of AFM state of AB bilayer CGT: (d) without an external electric field and (e) with an external electric field 0.8 V/Å, respectively. The isosurface is set $4\times10^{-12}$ e/bohr$^3$. Note that the positive and negative values correspond to the PDOS contributed by Cr/Te atoms in the bottom and top layers of CGT, respectively.

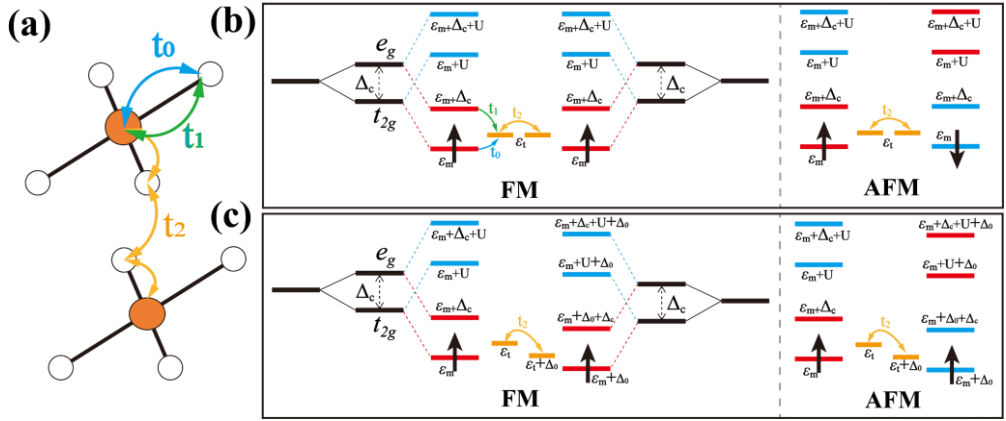

**Figure 3** (a) The schematic diagram of interlayer hopping model. The intralayer and interlayer hopping are represented by $t_0$, $t_1$ and $t_2$, respectively. (b) Schematic diagram of FM (left panel) and AFM (right panel) coupling for the interlayer hopping model without external electric field based on the mean-field approximation. (c) Schematic diagram of FM (left panel) and AFM (right panel) coupling for the interlayer hopping model with external electric field based on the mean-field approximation. The $\varepsilon_m$, $\varepsilon_t$, $\Delta_0$, $\Delta_C$ and U represent occupied Cr $t_{2g}$-orbital energy, Te p-orbital energy, on-site energy difference induced by external electric field, crystal filed energy and the strength of the Coulomb repulsion, respectively.

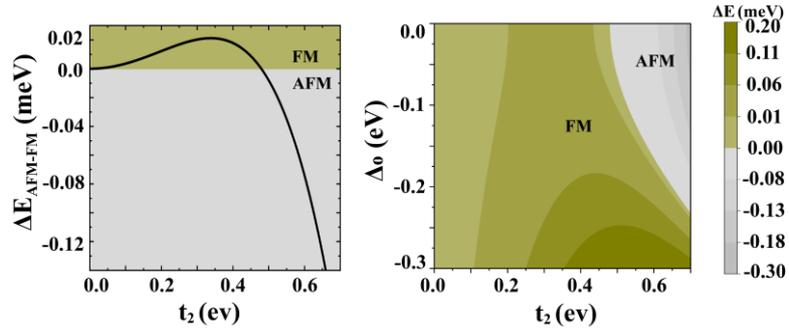

**Figure 4** (a) The exchange energy as a function of interlayer hopping parameter $t_2$. (b) The exchange energy as a function of band splitting $\Delta$ and interlayer hopping parameter $t_2$. The color-bar represents the exchange energy. According to the DFT calculations, the values of $\varepsilon_m$, $\varepsilon_t$, U and $\Delta_C$ are 0.0 eV, 0.80 eV, 3.50 eV and 2.50 eV are adopted, respectively. The $t_0$ and $t_1$ are set to 1.95 eV and 1.42 eV, respectively.